\documentclass[useAMS,usenatbib,usegraphicx]{mn2e}
\usepackage{amssymb,amsmath,amsfonts,graphicx}
\usepackage{epsfig,color}

\usepackage{pgf}
\usepackage{bm}

\def\gsim { \lower .75ex \hbox{$\sim$} \llap{\raise .27ex \hbox{$>$}} }
\def\lsim { \lower .75ex \hbox{$\sim$} \llap{\raise .27ex \hbox{$<$}} }

\usepackage{times}
\usepackage[bookmarks,bookmarksnumbered,colorlinks=true, citecolor=blue, linkcolor=black]{hyperref}
\usepackage{color}

\begin{document}

\title[The satellites of the Magellanic Clouds]{Identifying true satellites of the Magellanic Clouds}

\author[Sales et al.]{
\parbox[t]{\textwidth}{
Laura V. Sales$^{1}$$\thanks{E-mail: lsales@ucr.edu}$,     
Julio F. Navarro$^{2}$,
Nitya Kallivayalil$^{3}$
and Carlos S. Frenk$^{4}$     
}
\\
\\
$^{1}$ Department of Physics and Astronomy, University of California Riverside, 900 University Ave., CA92507, US\\
$^{2}$ Senior CIfAR Fellow. Department of Physics and Astronomy, University of Victoria, Victoria, BC V8P 5C2,
Canada\\
$^{3}$ Department of Astronomy, University of Virginia, Charlottesville, VA 22904,
USA\\
$^{4}$ Institute for Computational Cosmology, Department of Physics, University of Durham, South Road, Durham DH1 3LE, UK\\
}

\maketitle

\begin{abstract}
  The hierarchical nature of $\Lambda$CDM suggests that the Magellanic
  Clouds must have been surrounded by a number of satellites before
  their infall into the Milky Way. Many of those satellites
  should still be in close proximity to the Clouds, but some could
  have dispersed ahead/behind the Clouds along their Galactic
  orbit. Either way, prior association with the Clouds results in
  strong restrictions on the present-day positions and velocities of
  candidate Magellanic satellites: they must lie close to the
  nearly-polar orbital plane of the Magellanic stream, and their
  distances and radial velocities must follow the latitude dependence
  expected for a tidal stream with the Clouds at pericenter. We use a
  cosmological numerical simulation of the disruption of a massive subhalo in a
  Milky Way-sized $\Lambda$CDM halo to test whether any of the $20$
  dwarfs recently-discovered in the DES, SMASH, Pan-STARRS, and ATLAS
  surveys are truly associated with the Clouds. Of the $6$ systems
  with kinematic data, only Hydra~II and Hor~1 have distances and
  radial velocities consistent with a Magellanic origin. Of the
  remaining dwarfs, six (Hor~2, Eri~3, Ret~3, Tuc~4, Tuc~5, and Phx~2)
  have positions and distances consistent with a Magellanic origin,
  but kinematic data are needed to substantiate that
  possibility. Conclusive evidence for association would require
  proper motions to constrain the orbital angular momentum direction,
  which, for true Magellanic satellites, must coincide with that of
  the Clouds. We use this result to predict radial velocities and
  proper motions for all new dwarfs. Our results are relatively insensitive to the
  assumption of first or second pericenter for the Clouds.
\end{abstract}


\section{Introduction}
\label{sec:intro}

The Large and Small Magellanic Clouds (LMC and SMC, respectively) are
a galaxy pair orbiting together in the halo of the Milky Way and
provide a prime example of the nested hierarchy of structures expected
in the $\Lambda$CDM galaxy formation paradigm
\citep{Springel2008b}. Their physical association seems beyond doubt,
given their relative proximity, correlated kinematics, and abundant
evidence of past interaction \citep[for a recent review, see,
e.g.,][]{Donghia2015}.

The path of the Clouds around the Galaxy is well constrained by
precise estimates of their distances, positions, radial velocities and
proper motions, which indicate a nearly-polar orbit on a plane closely
aligned with the Magellanic Stream \citep{Kallivayalil2006}. The
Clouds are just past pericenter, since their Galactocentric radial
velocities are positive and much smaller than their tangential
velocities \citep[$V_t\sim 314$, $V_r\sim +64$ km/s for the LMC, see,
e.g.,][]{Kallivayalil2013}. Their orbit must also have a fairly large
apocentric radius, since their total speed ($|V_{\rm LMC}|\sim 321$
km/s) exceeds the circular velocity of the Milky Way ($\sim 220$ km/s)
by a substantial amount. A large apocenter implies a long orbital
period, which has led to the suggestion that the Clouds might be on
their first pericentric passage. 

This conclusion depends on the total mass assumed for the Milky Way
halo, as well as on its assumed outer radial profile
\citep{Besla2007}, but it would explain naturally why the LMC and SMC
are still so tightly bound. Indeed, if the Clouds were at first
pericenter then the Galactic tide would not have yet had time to
disrupt the pair nor to disperse fully the common (sub)halo they
inhabit. As a result, most other dwarf companions of the Clouds should
still lie in their close vicinity. Such ``Magellanic satellites'' have
long been speculated \citep[see, e.g.,][]{Lynden-Bell1995}, and their
existence would be consistent with the relatively common occurrence of
dwarf galaxy associations in the nearby Universe \citep{Tully2006}.
The immediate surroundings of the Clouds should thus be a fertile
ground to search for new dwarfs, as proposed by \citet[][S11
hereafter]{Sales2011}.

A full search for satellites around the Clouds would be extremely
valuable. One reason is that, in $\Lambda$CDM, the satellite
luminosity function is expected to be a nearly scale-free function
when expressed in units of the luminosity of the primary
\citep{Sales2013}. In other words, to first order, the Galactic
satellite abundance should be simply a scaled-up version of that of
the Clouds.  A complete catalogue of Magellanic faint and ultra-faint
satellites would be easier to compile (the relevant survey volume
is much smaller than the full Galactic halo) and could therefore
help to constrain the incompleteness of all-sky surveys of Galactic
satellites. In general, the surrounding of dwarf galaxies, especially
those in the field, are promising sites for the discovery of new faint
galaxies \citep{Sales2013,Wheeler2015}.   

A second application would be to clarify the effects of
environment on the star formation history of dwarfs
\citep{Donghia2008,Wetzel2015}. An unambiguous identification of
Magellanic origin would enable a direct comparison with Galactic
satellites of similar stellar mass that have evolved in a rather
different environment. Finally, Magellanic satellites might also
provide clues to the nature of dark matter: indeed, fewer 
satellites are expected around the Milky Way in general, and the LMC
in particular, if dark matter was ``warm'' rather than cold
\citep[see, e.g.,][]{Kennedy2014}.

Given this context, it is not surprising that the recent discovery of
a number of candidate dwarfs in southern surveys targeting the Clouds'
vicinity, such as the Dark Energy Survey \citep[DES;][]{Bechtol2015,
  Koposov2015, Drlica2015, Kim2015a, Kim2015b}, the Survey of the
MAgellanic Stellar History \citep[SMASH;][]{Martin2015}, as well as in
other large surveys, such as PAN-STARRS
\citep{Laevens2015}, and ATLAS \citep{Torrealba2016}, have attracted
much attention.  

While not all of these candidates have follow-up spectroscopy
confirming that they are dark matter-dominated dwarf galaxies rather
than star clusters---six have spectra thus far \citep{Walker2016,
  Kirby2015, Martin2016}---they do occupy the same region in the
size-luminosity plane as ultra-faint dwarf galaxies (M$_{\rm V}$
between -2.0 and -7.8 and half-light radii, $r_h$, between $\sim 18$
and $\sim1000$ pc). It is not clear either which of these dwarfs, if
any, have a Magellanic origin.

On that point, \citet{Deason2015} cite a statistical argument based on
abundance-matching models applied to massive subhalos in the ELVIS
simulations \citep{Garrison-Kimmel2014} to suggest that $2$-$4$ of the
$9$ then known DES candidates might have come into the Milky Way with
the LMC.  \citet{Yozin2015}, on the other hand, conclude, on the basis
of orbit models, that the majority of the DES dwarfs could have been
at least loosely associated with the Clouds. Yet another analysis
suggests, using tailor-made numerical simulations, that only about
half of the DES new dwarf galaxies are very likely to have been
associated with the LMC in the past \citep{Jethwa2016} .

Here, we take a complementary and targeted approach, using an LMC
analog subhalo identified in a fully cosmological simulation of a
Milky Way-sized halo in $\Lambda$CDM. We track the positions and
velocities of subhalo particles to constrain
the likely location in phase space of systems with prior association
with the Clouds. This is an extension of the analysis previously
presented in S11, who concluded that {\it none} of the
$26$ Milky Way satellites known at the time were convincingly
associated with the Clouds. The main goal of the present work is to
assess the likelihood of association with the Clouds of the
recently-discovered dwarfs, as well as to predict the radial
velocities and proper motions required for that association to be
true.

In \S~\ref{sec:sims} we describe our numerical set up, in
\S~\ref{sec:results} we present the main results, including the
expected sky distribution of the companion dwarfs, their radial
velocities, probability of association with the LMC, as well as their
3D orbits. We conclude
with a brief summary of our main conclusions in \S~\ref{sec:conc}.

\begin{center}
\begin{figure*}
\includegraphics[width=0.86\linewidth,clip]{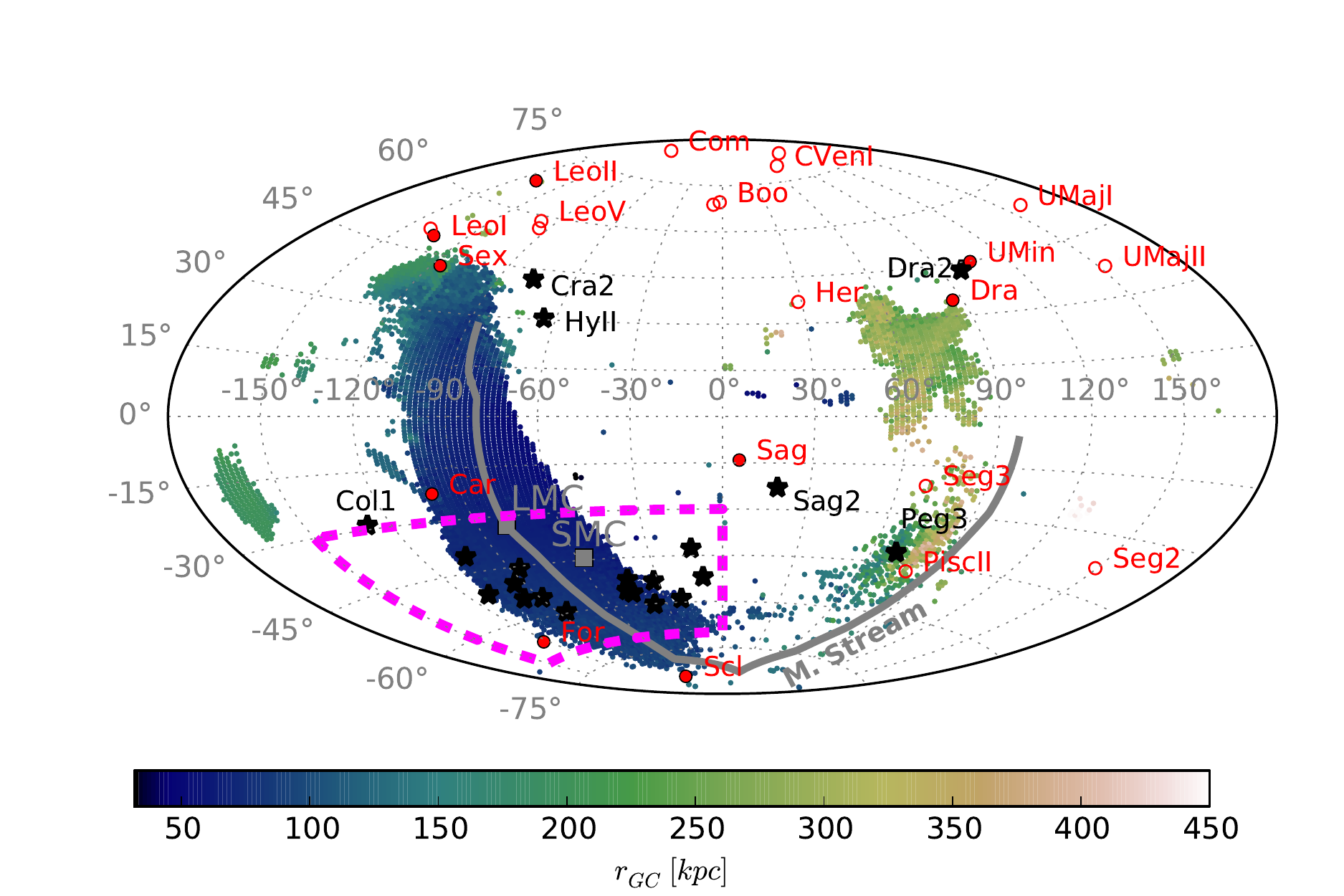}\\
\caption{Aitoff projection of particles associated with the LMC analog
  subhalo (LMCa), shown just after first pericentric
  approach, when its pericentric distance and velocity 
  closely matches that of the Large Magellanic Cloud. The LMCa center
  is chosen to coincide with the LMC and coordinates are chosen so
  that the direction of its orbital angular momentum matches that
  of the LMC. This results in a nearly-polar orbital pane, which
  roughly aligns with the Magellanic Stream (grey line).  Particles of
  the LMC analog (identified before infall) are colored by their average
  Galactocentric distance. Red circles indicate the position of known
  Milky Way satellites. Filled circles indicate ``classical'' dwarf
  spheroidals (i.e., brighter than $M_V=-8$); open circles denote
  fainter objects. Newly discovered dwarfs (the subject of this paper)
  are shown as black starred symbols.}
\label{fig:aitoff}
\end{figure*}
\end{center}
%
%

\section{Numerical Simulations}
\label{sec:sims}

We use the Aquarius Project \citep{Springel2008a}, a suite of
zoomed-in cosmological simulations that follow the formation of of $6$
Milky Way-sized halos with virial\footnote{We define the virial mass,
  $M_{200}$, as that enclosed by a sphere of mean density $200$ times
  the critical density of the Universe, $\rho_{\rm crit}=3H^2/8\pi
  G$.
  Virial quantities are defined at that radius, and are identified by
  a ``200'' subscript.} masses in the range
$0.8$-$1.8\; \times 10^{12} M_\odot$. These halos were selected from a
large scale simulation of a cosmologically representative volume
\citep[the Millennium-II Simulation, see][]{BoylanKolchin2009}. We
focus in this paper on the properties of an ``LMC analog'' system
(hereafter identified as LMCa, for short) which was identified and
presented in S11.

\subsection{LMCa: the LMC analog}

LMCa was chosen because it is a fairly massive subhalo with a
pericentric distance ($\sim 50$ kpc ) and velocity ($\sim 400$ km/s)
in good agreement with that of the LMC \citep[][hereafter
K06]{Kallivayalil2006}. Identified before infall, at $z_{id}=0.9$,
LMCa has a virial mass of $M_{200}=3.6 \times 10^{10} \; M_\odot$,
which corresponds to a circular velocity of $\sim 65$ km/s. 

LMCa first crosses the virial boundary of the main Aquarius halo (Aq-A)
at $z = 0.51$ ($t = 8.6$ Gyr), reaches first pericenter at $t_{\rm
  1p}=9.6$ Gyr, and is able to complete a second pericentric passage
at $t_{\rm 2p}=13.3$ Gyr. The host halo has a virial mass of
$M_{200}=1.8\times 10^{12}\, M_\odot$ at $z=0$. (These times are
actually slightly past actual pericenter, thus chosen so as to best
accommodate the fact that the LMC has a slight positive radial
velocity and is itself just past pericenter at present.)

At $t_{1p}$ and $t_{2p}$, the distances, radial velocities, and tangential
velocities are, respectively, $r_{1p}=65$ kpc, $r_{2p}=69$ kpc,
$V_{r,1p} = 78$ km/s, $V_{r,2p}=89$ km/s; $V_{t,1p} = 345$ km/s; and
$V_{t,2p}=302$ km/s. These values are in reasonable agreement with the
K06 LMC measurements (see Fig.~1 in S11), although the tangential
velocities were a bit below the observed values. The revised proper
motions for the LMC from \citet{Kallivayalil2013} suggest a slightly
lower total velocity than previously determined, $321 \pm 24$
km/s compared to $378 \pm 31$ km/s, resulting from the combination of
an added third-epoch of observations, the adoption of a different
local standard of rest, and a new determination of the LMC's dynamical
center. This decrease in velocity accommodates the tangential motion
of LMCa more comfortably at both pericenters.

It is still a matter of debate whether the Clouds are on first or
second pericentric passage \citep[see,
e.g.,][]{Shattow2009,Sales2011}, although indirect evidence favour a
first infall scenario, including: $a)$ their large tangential
velocity, $b)$ their blue colors and large gas content and $c)$ the
requirement that the LMC and SMC have been a long-lived binary (which
favors a low-mass Milky Way, or a high-mass LMC, see discussion in
\citealt{Kallivayalil2013}).  Therefore in what follows we analyze in
detail a first infall scenario but include a brief discussion about
how our conclusions would be affected if the LMC is in its second
pericenter passage (Sec.~\ref{sec:2nd_pericenter}).

Following S11, we use the Aquarius ``A'' halo at level 3 resolution,
or Aq-A-3 in the notation of \citet{Springel2008a}, which has a mass
per particle $m_p=4.9 \times 10^4 M_\odot$. We identify and follow all
particles that were associated with the LMCa friends-of-friends group
at the time of infall, and evaluate their positions and velocities at
the time of first and second pericenter passages.

Using {\sc subfind} \citep{Springel2001a}, we have identified more
than 200 subhalos associated with LMCa at infall time (see
Fig.~1 in S11 for their individual orbits), suggesting that a large
satellite such as the LMC should bring along its own population of
satellites \citep{Springel2008b}.  We use for our analysis {\it all}
particles (and not just the subhalos) initially bound to LMCa in order
to provide a more complete sampling of the positions and velocities of
any potential companion associated with the LMC.

\subsection{LMCa in Galactic coordinates}
\label{SecLMCaCoord}

We transform the coordinate system of the simulation into ``Galactic
coordinates'' by requiring that the orientation of the orbital angular momentum of LMCa
coincides with that measured for the LMC's orbit, and that its
position on the sky coincides with the LMC. For consistency with S11,
we use throughout this paper the LMC proper motion as given by
K06\footnote{We note however that the change in the direction of the
  orbit given by the new updated measurements from
  \citet{Kallivayalil2013} is very small:
  $(j_x,j_y,j_z)=(-0.97, 0.14, -0.18)$ versus $(-0.98, 0.11, -0.13)$
  for the 2006 and 2013 determinations, respectively. These numbers
  correspond to a unit vector in a Cartesian system aligned with the
  disk of the galaxy, as described in Sec.~\ref{sec:results}.}. After the
rotation, we also rescale sightly all Galactocentric distances so 
that LMCa is, at each pericenter, at the measured distance of the 
LMC: $49$ kpc.

%
\begin{center}
\begin{figure*}
\includegraphics[width=0.89\linewidth]{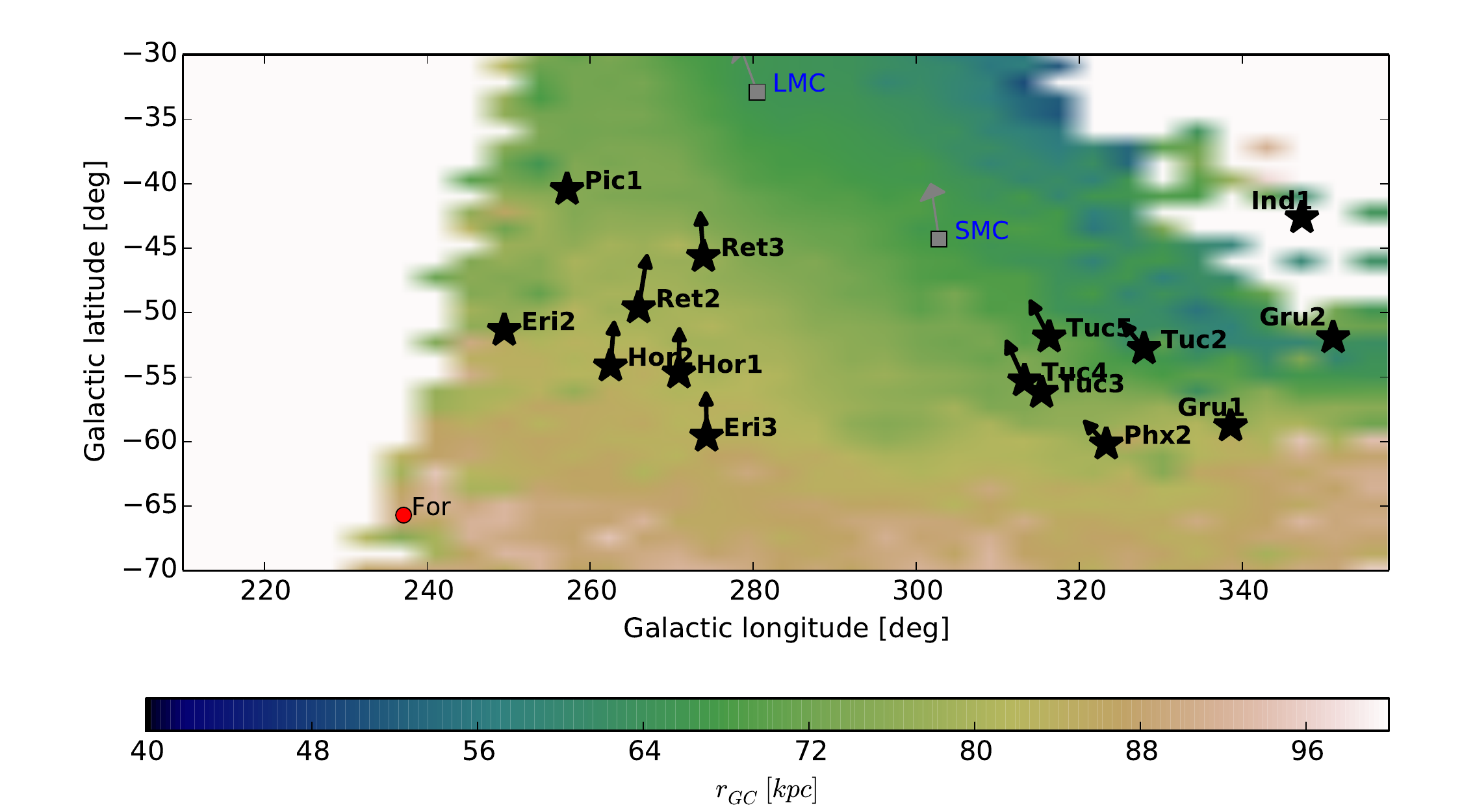}\\
\caption{Zoom-in of the area just south of the Clouds outlined by the
  dashed magenta box in Fig.~\ref{fig:aitoff}. This area samples the
  trailing arm of the LMCa tidal debris, and contains the new dwarfs
  discovered in the Dark Energy Survey (DES). Col~1 is the only DES
  dwarf located far away from the stream (not shown). Note also that
  Ind~1 has now been shown to be a star cluster \citep{Kim2015c}. The
  LMC and SMC are shown as grey squares; red circles are previously
  known Galactic satellites; new dwarfs are shown by starred
  symbols. The arrows indicate the expected tangential motion of those
  satellites, {\it assuming} that they were associated with the Clouds
  (see Sec.~\ref{SecPropMot}). Arrows are only shown for systems
  deemed likely Magellanic satellite candidates in a first pericenter
  passage scenario (see text for more
  details).}
\label{fig:aitoff_box}
\end{figure*}
\end{center}
%

\section{Results}
\label{sec:results}

We first examine the sky distribution of particles associated at
infall with the LMC analog subhalo (hereafter ``LMCa debris'', for
short). We use this footprint, as well as their radial and tangential
velocities, to compare with available data for the newly-discovered
dwarfs. As mentioned above, we shall interpret coincidence in sky
position, radial velocity and distance between debris particles and
observed dwarfs as evidence of a possible association with the LMC.

\subsection{LMCa debris: sky distribution and distances}
\label{SecLMCaSkyD}

At the time of the first pericenter, tidal disruption due to the host
halo has already set in, but most particles are still bound and close
to the subhalo center. The rest of the material is
distributed along a thick but well-defined tidal stream that follows
the projection of the subhalo's orbital path on the sky.  A leading
and trailing arm extend towards more positive and negative latitudes,
respectively. The distribution of this debris roughly agrees with the
position of the HI Magellanic Stream, sketched here by a line that
traces the high-density HI in the sky maps of \citet{Nidever2010}.

Most of the debris, however, is close to the current position of the
Clouds (grey squares indicate the observed positions of the LMC and
SMC). Particles are colour coded in Fig.~\ref{fig:aitoff} by their
Galactocentric distance (see color bar), which shows a clear gradient
along the stream with distances reaching up to $300$ kpc, well beyond the
virial radius of the main host.

For reference, we indicate the positions of all known Milky Way
satellites in the figure as well. Red filled circles correspond to the
``classical'' (i.e., brighter than $M_v=-8$) dwarf spheroidal (dSph)
companions of the Milky Way; open circles indicate the position of
previously known, fainter satellites. We refer the interested reader
to S11 for a discussion of the probability of association with the LMC
of those satellites.

The recently-discovered dwarfs that are the focus of this paper are
shown using black starred symbols in Fig.~\ref{fig:aitoff}.  We
include in this sample: (i) the dwarfs reported by \citet{Koposov2015}
from year-1 DES data \citep[see also][]{Bechtol2015}, (ii) the $6$
certain detections from year-2 DES data \citep{Drlica2015}, and (iii)
additional individual discoveries such as Hydra~II
\citep[][Hy~II]{Martin2015}, Horologium~2 \citep[][Hor~2]{Kim2015a},
Pegasus~3 \citep[][Peg~3]{Kim2015b}, Draco~2 and Sagittarius~2
\citep[][Dra~2 and Sag~2]{Laevens2015} and Crater~2
\citep[][Cra~2]{Torrealba2016}. Table~\ref{tab:prob} lists all the
``new dwarfs'' considered in what follows (i.e., black
stars in Fig.~\ref{fig:aitoff}). With the exception of Hy~II, Cra~2
and Dra~2, all other dwarfs are in the region of the sky occupied by
the trailing arm of the stream.

Fig.~\ref{fig:aitoff} shows that position on the sky and distance
provide on their own powerful constraints on a potential Magellanic
origin for a dwarf. Those satellites must be close to the orbital
plane (traced by the debris and the Magellanic Stream), ruling out
satellites like Sagittarius, Hercules, and Seg~2. In addition, the
farther a satellite is from the LMC the larger, on average, its
Galactocentric distance should be, a fact that rules out many of the
satellites in the Galactic northern cap. Indeed, the latter are
typically much closer to the Galactic centre than the leading arm of
the LMCa debris, which reaches a distance of $\sim 180$ kpc at
$b=+45^\circ$.

Fig.~\ref{fig:aitoff_box} zooms in on the vicinity of the LMC (the
region highlighted by the magenta box in Fig.~\ref{fig:aitoff}) and shows in
more detail the position of individual dwarfs as well as the distance
gradient expected for this section of the stream.  This figure also
shows that Col~1 lies outside of the LMCa debris footprint. This,
combined with its large distance ($\sim 182$ kpc) makes a Magellanic
association rather unlikely \citep[see also][]{Drlica2015}. We
therefore exclude Col~1 from the rest of our analysis, together with
Sag~2, whose position in the sky is not favorable either. Furthermore,
we also remove Indus 1 from our analysis since it has now been
classified as a stellar cluster rather than a dwarf galaxy
\citep{Kim2015c}.

%
\begin{center}
\begin{figure}
\includegraphics[width=100mm]{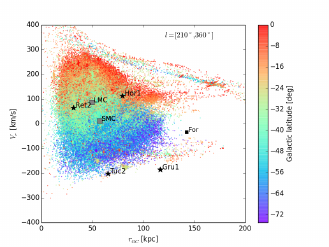}\\
\caption{Galactocentric distance $r_{GC}$ vs. radial velocity $V_r$
  for LMCa particles at first pericenter, color-coded by Galactic latitude $b$
  ($-80^\circ<b<0^\circ$; see color bar on right). For clarity, we
  only show the Galactic longitude range $l=[210^\circ,360^\circ]$,
  which encompasses most of the LMCa material in
  Fig.~\ref{fig:aitoff}.  Note the correlation
  between latitude and radial velocity, with the leading arm having
  already passed through pericenter (positive $V_r$) and the trailing
  material still approaching the Galaxy with $V_r < 0$. As before, the 
  LMC and SMC are shown with grey squares and other previously
  known dwarfs in this region of the sky are marked with black
  squares; new dwarfs with measured kinematics are shown with
  black starred symbols. The little overlap between Fornax, Gru~1 and
  Tuc~2 and the LMCa debris implies a low probability of prior
  association between these dwarfs and the LMC, assuming first
  infall. Hor~1 is the dwarf most likely to have
  had a Magellanic association. }
\label{fig:r_vr}
\end{figure}
\end{center}

The distance gradients with Galactic latitude shown in
Figs.~\ref{fig:aitoff} and ~\ref{fig:aitoff_box} result from the fact
that LMCa is close to pericenter and, therefore, at roughly  the minimum
distance of all associated debris.  Debris north of the LMC is
farther away and moving out (already past pericenter), whereas debris
to the south is also farther away but moving in (has yet to reach
pericenter). This induces a correlated signature in the radial
velocities, which we explore next.

\subsection{LMCa debris: radial velocities}
\label{SecLMCaRV}

We explore the correlation between radial velocity and Galactic
latitude in Fig.~\ref{fig:r_vr}. This figure shows the Galactocentric
radial velocity $V_r$ as a function of distance $r_{GC}$ for LMCa
debris in the Galactic longitude range $l=[210^\circ$$-$$360^\circ]$,
which encloses the stream and the positions of the DES dwarfs.

Particles are colored according to their Galactic latitude,
in the range $-80^\circ<b<0^\circ$ (see color
bar). Fig.~\ref{fig:r_vr} shows a clear gradient in radial velocity
with Galactic latitude, showing generally positive values (outward moving)
for particles north of the position of the LMC
(i.e., $b_{\rm LMC} > -32.9^\circ$) and negative values (infalling) for
those south of that. Although the latitude trend is clear, the
dispersion about the mean trend is quite large. This is 
because the velocity dispersion of LMCa before infall was quite
substantial (at $z_{\rm id}=0.9$, $\sigma_{200}=V_{200}/\sqrt 2=44.5$
km/s), making the tidally-induced stream quite thick. As a result, the
constraints on a possible Magellanic origin provided by $b$, $r_{\rm GC}$
and $V_r$ alone are relatively lax, and serve mainly to rule out the
most unlikely candidates.

For example, Fig.~\ref{fig:r_vr} shows that Fornax (even though it is
close to the stream in sky projection) has a distance that is too
large to be associated with the LMC, whereas the SMC, as expected,
lies well within the velocity-distance range spanned by the LMCa
debris. Starred symbols show the ``new dwarfs'' that fall in this
region of the sky  and for which kinematic measurements are available
\citep{Walker2016, Koposov2015b}: Hor~1, Ret~2 are clear candidates,
whereas Tuc~2 and Gru~1 seem only marginally consistent with a
Magellanic origin.

\begin{center}
\begin{figure}
\includegraphics[width=84mm]{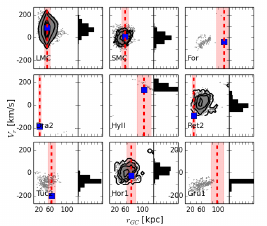}
\caption{Galactocentric distance vs. radial velocity 
  for LMCa particles within $5^\circ$ from each observed dwarf (blue
  squares), as labeled. Particles with $r_{GC}$ within $20\%$ of the observed
  distance fall within the red shaded area, and are used to
  ``predict'' the radial velocity expected for LMC association (see
  black velocity histograms on the right of each panel).  
  The top three panels are meant to illustrate
  the procedure for well studied systems. The LMC sits at the middle
  of the distribution {\it by construction}. The SMC is a likely LMC
  satellite; Fornax is not.  The bottom two rows show the newly
  discovered dwarfs for which kinematic measurements are
  available. Only Hy II and Hor 1 show velocities consistent with
  those expected for prior association with the LMC.  }
\label{fig:r_vr_each_wv}
\end{figure}
\end{center}

More stringent constraints may be obtained by combining the results
from Fig.~\ref{fig:aitoff} and Fig.~\ref{fig:r_vr}, since membership
to the LMC group is only likely for systems in narrow regions of the
four dimensional space drawn by (i) position on the sky $(l,b)$; (ii)
radial velocity $V_r$, and (iii) Galactocentric distance $r_{GC}$.  We
illustrate this in Fig.~\ref{fig:r_vr_each_wv}, where we plot the
distance and radial velocity of all LMCa particles whose positions on
the sky fall within $5^\circ$ of each individual dwarf.

The top two panels on the left of Fig.~\ref{fig:r_vr_each_wv} are
meant to illustrate the analysis procedure. For the case of the LMC
(top left) most particles in the LMCa subhalo are, by construction
(Sec.~\ref{SecLMCaCoord}), at the observed location and radial
velocity of the LMC (shown with a blue square). The SMC panel
illustrates that most LMCa particles selected in that direction of the
sky ($b=-44.3^\circ$, $l=302.8^\circ$) are at $\sim 58$ kpc from the
Galactic center and have, on average, a radial velocity of $\sim 5$
km/s, which is in excellent agreement with the observed SMC values
(blue square). 

The red vertical bands in the panels of Fig.~\ref{fig:r_vr_each_wv}
indicate a (generous) $20\%$ uncertainty in the distance estimate to
each dwarf; its intersection with LMCa particles is used to draw the
velocity histograms in the right-hand side of each panel.  Coincidence
between the velocity of the blue square and the histogram indicates
that the observed velocity is not unexpected in a scenario where the
dwarf originates from a disrupted LMC group. The velocity histograms
may therefore be used to ``predict'' the radial velocity of dwarfs for
which kinematic data is not yet available, {\it assuming} a Magellanic
origin.

As may be seen from Fig.~\ref{fig:r_vr_each_wv}, and not surprisingly,
the SMC passes these tests handily, making its association with the
LMC quite likely. On the other hand, the probability of association of
a dwarf like Fornax is quite remote. Most debris in that direction of
the sky are at much closer distances, and the little that overlaps in
distance with Fornax (two particles) has a rather high positive radial
velocity, quite unlike that observed. This illustrates the arguments
used by S11 to exclude an LMC association not only for Fornax but also
for all other Galactic satellites known at that time in case of first infall.

The $6$ ``new dwarfs'' with kinematic data are shown in the bottom two
rows of Fig.~\ref{fig:r_vr_each_wv}. From this comparison we conclude
that Dra~2 has little chance of LMC association. Likewise, Ret~2,
Tuc~2 and Gru~1 have velocities only marginally consistent with a
Magellanic relation. Hy~II, on the other hand, has the correct radial
velocity for its distance, despite its large angular separation from
the LMC, at the far northern edge region of the leading stream. The
only clear candidate for Magellanic association is Hor~1, which is
well within the expected velocity-distance range at its location.

%

\begin{center}
\begin{figure}
\includegraphics[width=0.89\linewidth,clip]{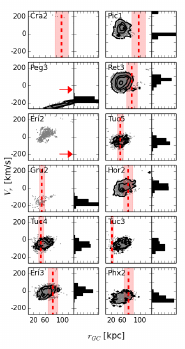}
\caption{Same as Fig.~\ref{fig:r_vr_each_wv} but for the new dwarfs
  with no measured $V_r$. The Galactocentric distances for
  Pic~1,  Eri~2 and Tuc~3 seem inconsistent with the distances
  measured for the LMCa debris around their positions on the
  sky. On the other hand, Ret~3, Hor~2 and several of the Tucanas show high
  chance of association, at least based on their positions and distances alone.}
\label{fig:r_vr_each_nov}
\end{figure}
\end{center}
%

\subsection{Predicted radial velocities for candidate Magellanic satellites}
\label{SecMagSatRV}

We can use the procedure described in the previous subsection to
predict the radial velocities that the remaining ``new dwarfs'' would
have if they were truly Magellanic satellites. We show this in
Fig.~\ref{fig:r_vr_each_nov}, which lists dwarfs in order of
decreasing Galactic latitude. Inspection of individual panels suggests
some preliminary conclusions. The Galactocentric distances measured
for Pic~1 and Eri~2 seem inconsistent with previous association with
the LMC. Cra~2 is in the same category, given the very little overlap
with the edge of the leading arm of the stream.  Aside from those
three cases, all other dwarfs show some degree of overlap in the
$(l,b)$-$r_{GC}$ plane with the LMCa debris. For the latter, the black
histograms in Fig.~\ref{fig:r_vr_each_nov} show their expected radial
velocities for a Magellanic origin. We summarize these predictions in
Table~\ref{TabPropMot2}, together with uncertainties derived from the
interquartile velocity range of the histograms in
Fig. \ref{fig:r_vr_each_nov}.

\subsection{Probability of LMC association}
\label{SecLMCProb}

The discussion of Figs.  ~\ref{fig:r_vr_each_wv} and
~\ref{fig:r_vr_each_nov} suggests that, for each dwarf, the
probability of prior LMC association scales with the total number of
LMCa particles that match its sky position, distance, and radial
velocity. We emphasize that these are not probabilities in the
statistical ``likelihood'' sense, but nevertheless provide a simple
way to rank order the dwarfs in terms of their potential association
with the LMC and to weed out unrelated systems.

We adopt the following procedure, which attempts to quantify how
likely the observed position (and velocity, when available) of a dwarf
is, assuming LMCa membership. To this aim, we first compute, for each
LMCa particle, the radius of a sphere, $r_{100p}$, that contains the
$100$ nearest debris particles, and rank them by this
metric\footnote{We have checked that none of our conclusions change
  when selecting, instead, $50$ or $200$ particles for this
  exercise.}. The smaller $r_{100p}$ the more closely associated a
particle is to the stream, which suggests that we may use $r_{100p}$
to define a probability of association. In other words, we assign a
dwarf a ``probability of association'' equal to the fraction of LMCa
particles with $r_{100p}$ values greater than that computed using the
dwarf's position. Probabilities assigned in this manner are listed in
Table~\ref{tab:prob} for all newly-discovered dwarfs'.

For dwarfs with measured radial velocities, we compute a further
probability by comparing its radial velocity with that of the nearest
$100$ LMCa particles. In practice, we use the mean and dispersion of
those $100$ radial velocities to compute the probability that the
observed velocity of the dwarf was drawn at random from that
distribution, assuming Gaussian statistics. The probabilities listed
in columns~10 and ~11 of Table~\ref{tab:prob} are computed by
multiplying this value by that estimated using the position alone
(columns~8 and ~9, respectively).

The results are shown in Table~\ref{tab:prob}, where we list all ``new
dwarfs'', as well as their assigned probability, with and without
velocity information.  As discussed before, aside from the SMC, the
procedure ranks Hor~1 as the best candidate for a true Magellanic
satellite when considering satellites with or without radial
velocities. Of systems without kinematic data, Hor~2, Eri~3 and Ret~3
are the best candidates, but this could certainly change when radial
velocities become available. An example of the importance of kinematic
information is provided by Ret~2, whose probability drops
substantially (from $\sim 0.10$ to $0.01$) when adding its radial velocity
to the analysis (assuming first pericenter).

We shall hereafter retain as ``Magellanic candidates'' systems whose
probabilities exceed $50\%$ that obtained for the SMC, without
velocity information. The list of candidates is quite short: only $7$
systems of the $20$ new dwarfs make the cut in the case of
position-only information: Hor~1, Hor~2, Eri~3, Ret~3, Tuc~5, Tuc~4
and Phx~2.

\subsection{Second pericenter}
\label{sec:2nd_pericenter}

The above  procedure also allows us to explore the sensitivity of our
findings to our assumption that the LMC is on first approach. We do
this by performing the same analysis but using the LMCa data at second
pericenter ($t_{2p}$), after updating the Galactic coordinate system
transformation described in Sec.~\ref{SecLMCaCoord}. The new
probabilities are also listed in Table~\ref{tab:prob}. 

Because of the procedure, probabilities are nominally higher, on
average, at second pericenter. This is because the LMCa debris spreads
out further in phase space at second pericenter, thus generally
boosting the probability values computed for most systems. In general,
however, there is a strong correlation between the probabilities at
both pericenters, so our conclusions seem only weakly dependent on the
assumption of first infall.

The increase in probability is most notable in the cases of Dra~2 and
Hy~II, whose probabilities jump from $\sim 0.01$ and $0.01$ in a first
pericenter passage to $0.50$, and $0.16$, respectively, when
considering the second pericentric passage. The jump in probability is
even more remarkable when including velocities, reaching $0.47$ in the
case of Dra~2 at second pericenter, more than for the SMC.

The reason for this, in the case of Dra~2, is that it sits at the very
far edge of the ``trailing arm'' of the tidal stream. Although its
distance and velocity are consistent with an LMC association, at first
infall there are only a few particles at that sky location and its
probability is quite low. When the Clouds are in a second passage,
several particles accumulate near the apocenter of the LMCa orbit, not
far from where Dra~2 is, increasing substantially its probability of
association. Similarly, Hy~II is at the tip of the leading arm of the
stream, a position that is much more heavily populated after the
Clouds have completed one full orbit around the Galaxy.

\subsection{Proper motions: conclusive proof of Magellanic origin}
\label{SecPropMot}

Ultimately, the most compelling evidence for LMC association will come
from the proper motions of the new dwarfs. This is because all
material associated with the LMC before infall is expected to retain
the direction of its orbital angular momentum. In other words, to
first order, Galactic tides are not expected to torque the LMC or its
debris away from their original orbital plane.

\begin{center}
\begin{figure}
\includegraphics[width=84mm]{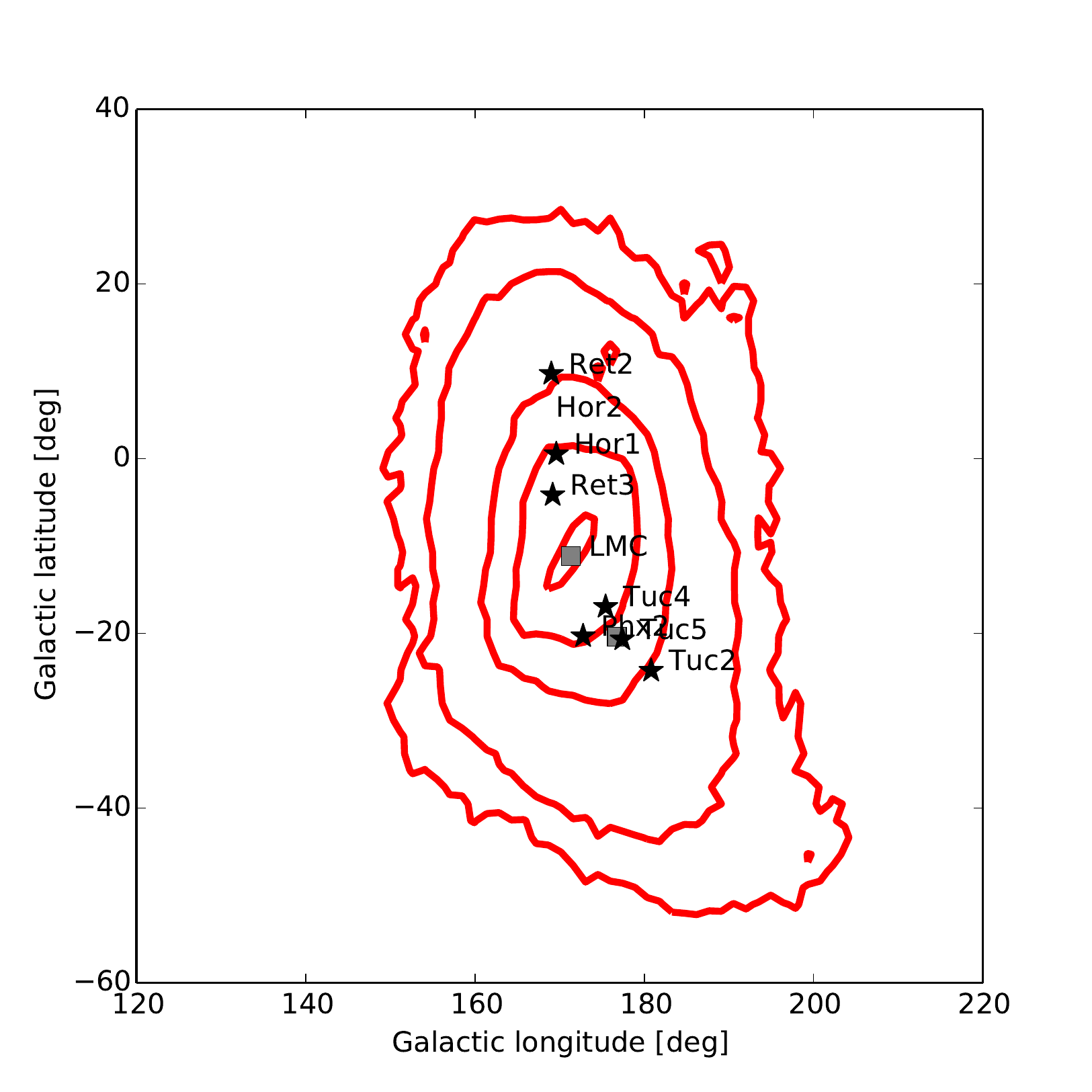}
\caption{Sky coordinates of the orbital poles (i.e., the direction of
  the orbital angular momentum) of particles associated with
  LMCa. Contours show constant density lines for the distribution of
  all LMCa.  Symbols correspond to the orbital poles estimated for new
  dwarf galaxies deemed likely candidate Magellanic satellites at
  first pericenter, as labeled. These estimates are based on the
  particles in the stream that are close in the sky and that lie at
  distances within $20\%$ of the measured values (see shaded red
  regions in Fig.~\ref{fig:r_vr_each_wv} and
  \ref{fig:r_vr_each_nov}). The common infall scenario preserves the
  coherence of the orbital plane, resulting in a tight distribution of
  the orbital poles in a well-defined region of the sky.}
\label{fig:poles}
\end{figure}
\end{center}

We show this in Fig.~\ref{fig:poles}, where we plot the {\it direction} of the
orbital angular momentum of LMCa particles at first
pericenter. The innermost and outermost isodensity contours enclose $5\%$,
and $95\%$ of all LMCa particles, respectively, and are centered at
the location of the LMC orbital pole (central grey
square). The other grey square (at $b=-20^\circ$ and $l=175^\circ$)
corresponds to the SMC, and is consistent with its assumed
association with the LMC.

We also show with starred symbols in Fig.~\ref{fig:poles} the orbital
angular momentum direction predicted for each of the ``candidate
Magellanic satellites'' (i.e., those that exceed $50\%$ probability
compared to the SMC) at first pericenter {\it assuming} that they were
associated with the LMC. This is computed as the median $l$ and $b$ of
the orbital poles of all LMCa particles with matching sky position and
Galactocentric distance (i.e., the particles that fall
into the red shaded regions of each panel in
Figs.~\ref{fig:r_vr_each_wv} and \ref{fig:r_vr_each_nov}).

We list in Table~\ref{TabPropMot1}, for each Magellanic candidate, the
coordinates of the predicted orbital angular momentum unit vector, in
a Cartesian system where the $Z$-axis is perpendicular to the Galactic
disk, the $X$-axis points away from the sun and the $Y$-axis is
defined such that we get a right-handed system.  Uncertainties
correspond to the r.m.s. values from the individual LMCa particles
used for each dwarf.

Assuming a radial velocity (for those without a measurement), this is
equivalent to predicting the tangential motion of each dwarf, which we
also list in Table~\ref{TabPropMot2}. The predicted projected
velocities are shown with arrows in Fig.~\ref{fig:aitoff_box}.
Table~\ref{TabPropMot1} may therefore be used to evaluate the
hypothesis of prior LMC association for these dwarfs once proper
motions for these objects become available.

\section{Summary}
\label{sec:conc}

We have used a $\Lambda$CDM cosmological N-body simulation of the
formation of a Milky Way-sized halo to investigate which of the $20$
newly-discovered Galactic satellites in the DES, Pan-STARRS, SMASH and
ATLAS surveys might have been associated with the Magellanic Clouds
before infall. Our study extends that of \citet{Sales2011}, which used
a massive subhalo with orbital parameters that closely match those of
the LMC (an LMC analog: LMCa) and tracked the position and velocity of
its constituent particles at first and second pericentric
passages. This enables the probability of LMC association to be
assessed by checking whether individual dwarfs lie in a region of
phase space populated by debris from the disrupting LMC subhalo.

On the basis of that analysis, \citet{Sales2011} concluded that,
except for the SMC, none of the other $26$ Galactic satellites known
at the time had positions and velocities consistent with a Magellanic
origin. That was the first study to investigate a possible Magellanic
association by using all of the available phase-space information in a
fully cosmological context. Extending this study to the
newly-discovered dwarfs yields the following conclusions, assuming
that the LMC is at first pericentric passage.

\begin{itemize}

\item We eliminate four systems from our analysis. Sag~2, Eri~2,
  and Col~1 lie too far outside the LMCa footprint for their
  association with the LMC to be plausible. In addition, Ind~1 has recently
  been reclassified as a star cluster.  

\item For the rest of the systems, a quantitative ``probability'' of
  association has been computed using the positions and velocities of
  the LMCa particles closest to each dwarf. We deemed likely candidate
  Magellanic satellites dwarfs whose probability exceeds half the
  value assigned to the SMC.

\item Of the six systems with available distances and radial
  velocities, only Hor~1 is clearly consistent with a Magellanic
  origin. Ret~2, Tuc~2, and Gru~1 have radial velocities which are
  only marginally consistent with LMC association. Dra~2 is too far
  off the LMCa first-pericenter footprint. Hy~II has the right distance
  {\it and} radial velocity, but its probability is small, given its
  position at the thinly-populated, very far end of the LMCa leading
  tidal arm.

\item Of the remaining $11$ systems with only sky positions and
  distances, our analysis retains $6$ of them at higher than
  $50\%$ the probability of the SMC (Hor~2, Eri~3, Ret~3, Tuc~5, Tuc~4, and
  Phx~2). For these candidates, the nearest LMCa particles are used to
  {\it predict} their radial velocities, assuming a
  Magellanic origin.

\item Aside from radial velocities, the most telling evidence of a
  potential LMC association would be provided by proper motions. These
  constrain the direction of the orbital angular momentum of each
  dwarf, which must roughly coincide with that of the LMC. We use this
  result to predict proper motions for all newly-discovered
  satellites, again assuming a Magellanic origin. The radial and
  tangential velocity predictions could be used to reassess the
  hypothesis of a possible Magellanic association once kinematic data
  become available.

\end{itemize}

Our conclusions are insensitive to our choice of first or second
pericenter for the LMC, in the sense that the association
probabilities of most dwarfs computed at each time show strong
correlation. Because the LMCa debris spreads out to cover a larger
volume in phase-space at second pericenter, the probabilities of four
extra systems, computed using positions alone, are lifted above
$50\%$ that of the SMC: Tuc~2, Dra~2, Cra~2, and Peg~3. Of these, Tuc~2 seems quite
unlikely given its radial velocity.  Dra~2, on the other hand, has
position and velocity consistent with being at the far end of the
trailing stream during a second pericenter.

Our main conclusion is therefore that few of the newly discovered
dwarfs are definitely associated with the LMC.  This is not entirely
unexpected. The simple scaling argument of \citet{Sales2013} suggests
that the fraction of all Galactic satellites associated with the
Clouds should be close to the ratio of the stellar mass of the LMC and
the Milky Way, i.e., $~\sim 5\%$. Given that we now have identified a
total $\sim 46$ dwarfs within $300$ kpc from the Galactic center
(excluding the LMC/SMC pair), only $2$ to $3$ should, in principle, be
associated with the Clouds. So far our analysis seems consistent with this
expectation. Accurate radial velocities and proper motions are needed
to accept/reject the hypothesis of association between these dwarfs
and the LMC. Confirming the existence of multiple Magellanic satellites
would provide a wonderful confirmation of the hierarchical nature of
galaxy formation predicted by the current cosmological paradigm.

\section{Acknowledgments}

NK is supported by the NSF CAREER award 1455260. This research was
supported in part by the National Science Foundation under Grant
No. NSF PHY11-25915 and by the hospitality of the Kavli Institute for
Theoretical Physics at the University of California, Santa Barbara.

\begin{center}
\begin{table*} 
  \caption{Parameters of the newly-discovered dwarfs considered in
    this paper, together with their
    probability of association with the LMC in either first or second
    pericenter passage, as defined in
    Sec.~\ref{SecLMCProb}. Cols.~8 and ~9
    list probabilities computed using positions alone; cols.~10 and
    ~11 also include also radial velocity data. We list the $V$-band absolute magnitude, 
    stellar mass, galactocentric coordinates $(l,b)$, measured
    heliocentric velocity $V_\odot$ and heliocentric distance 
    $D_\odot$ of each satellite, taken from the
    following references: [1]
    \citealt{Koposov2015}, [2] \citealt{Drlica2015}, [3]
    \citealt{Martin2015}, [4] \citealt{Laevens2015}, [5]
    \citealt{Kim2015a}, [6] \citealt{Kim2015b}, [7]
    \citealt{Torrealba2016}, [8] \citealt{McConnachie2012}, [9] \citealt{Simon2015},
    [10] \citealt{Koposov2015b}, [11] \citealt{Kirby2015}, 
    [12] \citealt{Walker2016},[13] \citealt{Martin2016}). For cases
    without estimates of $M_*$ we derive it from their listed V-band
    magnitudes assuming a mass-to-light ratio $\gamma=2$ in solar
    units. Dwarfs are grouped according to their probability of association
    with the Clouds at first pericenter, using only their distance and 
    position on the sky (col.~8). The main two groups include  ``likely
    candidates'' and ``unlikely candidates'', according to whether
    their probabilities are above or below $50\%$ the probability
    assigned to the SMC. The final group
    lists those that were discarded from the analysis, either because
    their position on the sky is such that the probability of LMC
    association is remote, or because they are considered star
    clusters, and not dwarf galaxies.}
\begin{tabular}[width=0.85\linewidth,clip]{|c|c|c|c|c|c|c|c|c|c|c|c|}
\hline
Name & $M_V$ & $M_{*}$ & $l$ & $b$ & $V_\odot$ & $D_\odot$ &
Prob$_{1st\; \rm per}$ & Prob$_{2nd\; \rm per}$ & Prob$_{1st\; \rm per}$ & Prob$_{2nd\; \rm per}$ & Refs.\\
 & [mag] & $[10^3 \rm M_\odot]$ & [deg] & [deg] & [km/s] & [kpc] & ($l$,$b$,$r$) & ($l$,$b$,$r$)  & ($l$,$b$,$r$,$V_r$)  & ($l$,$b$,$r$,$V_r$)  &\\
\hline
\hline
LMC & $-18.1$ & $1.5 \times 10^6$ & $280.5$ & $-32.9$ & $262.2$ & $51$  & $0.63$ & $0.71$ & $0.52$ & $0.53$ & [8]\\ 
SMC & $-16.8$ & $4.6 \times 10^5$ & $302.8$ & $-44.3$ & $145.6$ & $64$  & $0.28$ & $0.65$ & $0.21$ & $0.39$ & [8]\\ 
\hline
Hor~1 & $-3.4$ & $1.96$ & $270.9$ & $-54.7$ & $112.8 \pm 2.5$ & $79$ & $0.30$ & $0.66$  & $0.17$ & $0.20$ & [1],[10]\\ 
Hor~2 & $-2.6$ & $2.47$ & $262.5$ & $-54.1$ & & $26$ & $0.27$ & $0.59$  & &  & [5]\\ 
Eri~3 & $-2.0$ & $0.54$ & $274.3$ & $-59.6$ & & $87$ & $0.25$ & $0.64$ &  & &  [1] \\ 
Ret~3 & $-3.3$ & $13.0$ & $273.9$ & $-45.6$ & & $92$ & $0.25$ & $0.66$  & &  & [2] \\ 
Tuc~5 & $-1.6$ & $9.$ & $316.3$  & $-51.9$ & & $55$ & $0.15$ & $0.46$ &  &  &  [2] \\ 
Tuc~4 & $-3.5$ & $4.$ & $313.3$ & $-55.3$ & & $48$ & $0.15$ & $0.35$  & &  &  [2] \\ 
Phx~2 & $-2.8$ & $1.13$ & $323.3$ & $-60.2$ & & $83$ & $0.15$ & $0.55$ &  & &  [1] \\ 
\hline
Tuc~2 & $-4.4$ & $4.9$ & $327.9$ & $-52.8$ & $-129.1 \pm 3.5$ & $69$ & $0.11$ & $0.47$ &  $0.06$ & $0.06$ &  [1],[12] \\ 
Gru~2 & $-3.9$ & $5.0$ & $351.1$ & $-51.9$ & & $53$ & $0.09$ & $0.17$ &  & &  [2] \\ 
Ret~2 & $-2.7$ & $1.0$ & $265.9$ & $-49.6$ & $62.8 \pm 0.5$ & $30$ & $0.09$ & $0.10$  & $0.01$ & $0.03$ &  [1],[9],[10] \\ 
Tuc~3 & $-2.4$ & $2.0$ & $315.4$ & $-56.2$ & & $25$ & $0.08$ & $0.10$  & & &  [2] \\ 
Gru~1 & $-3.4$ & $1.96$ & $338.6$ & $-58.8$ & $-140.5 \pm 2.0$ & $120$ & $0.06$ & $0.17$ &  $<0.01$ & $0.01$ &  [1],[12] \\ 
Pic~1 & $-3.1$ & $1.5$ & $257.1$ & $-40.4$ & & $114$ & $0.06$ & $0.09$ &  &  &  [1] \\ 
Peg~3 & $-4.1$ & $7.46$ & $69.8$ & $-41.8$ & & $26$ & $0.05$ & $0.32$  & &  &  [6] \\ 
Cra~2 & $-8.2$ & $2.25$ & $283.9$ & $+41.9$ & & $118$ & $0.02$ & $0.41$ & & & [7]   \\ 
Dra~2 & $-2.9$ & $2.47$ & $98.3$ & $+42.9$ & $-347.6 \pm 1.8$ & $20$ & $0.01$ & $0.50$ & $0$ & $0.47$ &  [4],[13] \\ 
Hy~II & $-4.8$ & $7.1$ & $295.6$ & $+30.5$ & $303.1 \pm 1.4$ & $134$ & $0.01$ & $0.16$ & $0.01$ & $0.12$ &  [3],[11] \\ 
\hline
Eri~2** & $-6.6$ & $37.3$ & $249.4$ & $51.4$ & & $380$ & $0.0$ & $0$ &  &  &  [1] \\ 
Sag~2** & $-5.2$ & $2.47$ & $189.0$ & $-22.9$ & & $67$ & -- & --  & &  &  [4] \\ 
Ind~1** & $-3.5$ & $2.1$  & $347.3$ & $-42.6$ & & $100$ &  -- & --  & &   & [1] \\ 
Col~1** & $-4.5$ & $18.0$ & $231.6$ & $-28.9$ & & $182$ & -- & --  & &  &  [2] \\ 
\hline
\hline
\end{tabular}
\label{tab:prob}
\end{table*}
\end{center}
%

\begin{center}
\begin{table*} 
  \caption{Cartesian components of the direction (average) of the
    angular momentum of the LMCa particles near each Magellanic candidate dwarf,
    according to the discussion of  Sec.~\ref{SecLMCProb}. All vectors are normalized to have
    modulus unity. For each dwarf, we list the results for the first
    (top row) and/or second (bottom row) pericenter passage. The
    bottom group includes dwarfs that are only likely Magellanic candidates at second pericenter.  Because the
    LMC is in a nearly polar orbit, the angular momentum of all
    material associated with it points in all cases in the $-X$
    direction (i.e., to the Sun from the Galactic center). }
\begin{tabular}[width=0.85\linewidth,clip]{|c|c|c|c|c|}
\hline
Name & time &  $j_X$ & $j_Y$ & $j_Z$ \\
\hline
\hline
LMC & $t_{1p}$ & $-0.97 \pm 0.03$ &  $0.14 \pm 0.07$ & $-0.19 \pm 0.10$\\ 
    & $t_{2p}$ & $-0.97 \pm  0.03$ & $0.14 \pm 0.06$ &  $-0.18 \pm 0.09$\\ 
    & observed     & $-0.97 \pm 0.01$ & $0.14 \pm 0.02$  &  $-0.18 \pm 0.03$\\
SMC & $t_{1p}$ & $-0.92 \pm 0.05$ & $0.04 \pm 0.10$ & $-0.35 \pm 0.08$ \\ 
    & $t_{2p}$ & $-0.90 \pm 0.05$ & $0.05 \pm 0.17$ & $-0.38 \pm 0.10$  \\ 
    & observed & $-0.91 \pm 0.05$ & $0.08 \pm 0.11$ & $-0.39 \pm 0.09$ \\
\hline
Hor~1 & $t_{1p}$ & $-0.98 \pm 0.05$ & $0.18 \pm 0.10$ & $-0.04 \pm 0.09$ \\ 
     & $t_{2p}$ & $-0.95 \pm 0.19$ & $0.30 \pm 0.50$ & $-0.10 \pm 0.36$ \\ 
Hor~2 & $t_{1p}$ & $-0.97 \pm 0.02$ & $0.24 \pm 0.09$ & $-0.03 \pm 0.08$ \\ 
     & $t_{2p}$ & $-0.73 \pm 0.18$ & $-0.48 \pm 0.46$ & $0.49 \pm 0.28$ \\ 
Eri~3 & $t_{1p}$ & $-0.99 \pm 0.07$ & $0.16 \pm 0.10$ & $-0.04 \pm 0.07$ \\ 
     & $t_{2p}$ & $-0.94 \pm 0.20$ & $0.31 \pm 0.61$ & $-0.14 \pm 0.37$ \\ 
Ret~3 & $t_{1p}$ & $-0.98 \pm 0.02$ & $0.18 \pm 0.07$ & $-0.11 \pm 0.08$ \\ 
     & $t_{2p}$ & $-0.94 \pm 0.19$ & $0.28 \pm 0.50$ & $-0.15 \pm 0.44$ \\ 
Tuc~5 & $t_{1p}$ & $-0.93 \pm 0.03$ & $0.12 \pm 0.13$ & $-0.34 \pm 0.05$ \\ 
     & $t_{2p}$ & $-0.90 \pm 0.04$ & $0.09 \pm 0.14$ & $-0.42 \pm 0.08$ \\ 
Tuc~4 & $t_{1p}$ & $-0.95 \pm 0.03$ & $-0.06 \pm 0.17$ & $-0.30 \pm 0.06$ \\ 
     & $t_{2p}$ & $-0.93 \pm 0.03$ & $0.13 \pm 0.12$ & $-0.28 \pm 0.08$ \\ 
Phx~2 & $t_{1p}$ & $-0.93 \pm 0.02$ & $0.13 \pm 0.10$ & $-0.34 \pm 0.04$ \\ 
     & $t_{2p}$ & $-0.92 \pm 0.02$ & $0.13 \pm 0.13$ & $-0.37 \pm 0.05$ \\ 
\hline
Tuc~2   & $t_{2p}$ & $-0.87 \pm 0.03$ & $0.08 \pm 0.16$ & $-0.49 \pm 0.06$ \\ 
Peg~3     & $t_{2p}$ & $-0.97 \pm 0.02$ & $0.25 \pm 0.17$ & $-0.03 \pm 0.12$ \\ 
Cra~2    & $t_{2p}$ & $-0.97 \pm 0.08$ & $0.12 \pm 0.17$ & $0.20 \pm 0.16$ \\ 
Dra~2    & $t_{2p}$ & $-0.81 \pm 0.08$ & $-0.56 \pm 0.19$ & $0.19 \pm 0.20$ \\ 
%
\hline
\hline
\end{tabular}
\label{TabPropMot1}
\end{table*}
\end{center}

\begin{center}
\begin{table*} 
  \caption{Predicted Galactocentric radial and tangential velocity for
    Magellanic candidate dwarfs under the assumption
  of association with the Clouds. We show the median and $25\%$-$75\%$ percentiles in the case of first 
  (columns 2-4) and second (columns 5-7) pericenter passage. The last column shows the galactocentric 
  radial velocity for the 6 dwarfs with measured kinematics. The
    bottom group includes dwarfs that are only likely Magellanic
    candidates at second pericenter.}
\begin{tabular}[width=0.85\linewidth,clip]{|c|c|c|c|c|c|c|c|}
\hline
Name & $V_r^{\rm pred}$ & $V_l^{\rm pred}$  & $V_b^{\rm pred}$ & $V_r^{\rm pred}$ & $V_l^{\rm pred}$ & $V_b^{\rm pred}$ & $V_r^{\rm obs}$ \\
 & [km/s] & 1st per. & 1st per. & 2nd per. & 2nd per. & 2nd per.  &[km/s]\\
\hline
\hline
Hor~1 & $19^{-22}_{+23}$ & $5^{-30}_{+27}$ & $330^{-25}_{+17}$ & $23^{-17}_{+22}$  & $1^{-38}_{+73}$ & $266^{-200}_{+30}$ & $-23.2$ \\  
Hor~2 & $20^{-10}_{+25}$ & $44^{-22}_{+26}$ & $326^{-17}_{+19}$ & $30^{-26}_{+36}$  & $70^{-36}_{+13}$ & $56^{-21}_{+216}$ & \\  
Eri~3 & $5^{-23}_{+23}$ & $-57^{-26}_{+78}$ & $323^{-26}_{+22}$ & $22^{-39}_{+31}$ & $8^{-45}_{+65}$ & $252^{-200}_{+28}$ &   \\  
Ret~3 & $72^{-32}_{+16}$ & $36^{-15}_{+21}$ & $326^{-19}_{+25}$ & $38^{-23}_{+26}$  & $-8^{-44}_{+86}$ & $244^{-180}_{+35}$ &  \\ 
Tuc~5 & $-58^{-21}_{+26}$ & $-225^{-24}_{+27}$ & $272^{-28}_{+23}$ & $-50^{-27}_{+28}$ & $-245^{-17}_{+30}$ & $221^{-21}_{+32}$ &   \\ 
Tuc~4 & $-58^{-21}_{+24}$ & $-213^{-25}_{+26}$ & $297^{-25}_{+23}$ & $-43^{-28}_{+18}$  & $-221^{-24}_{+28}$  & $256^{-24}_{+24}$ &   \\ 
Phx~2 & $-90^{-6}_{+30} $ & $-273^{-8}_{+34}$ & $180^{-5}_{+30}$ & $-58^{-24}_{+21}$ & $-237^{-18}_{+18}$  & $161^{-24}_{+25}$ &  \\  
\hline
Tuc~2 & -- & -- & -- & $-73^{-19}_{+29}$ &  $-265^{-18}_{+23}$ & $133^{-22}_{+33}$ & $-201.5$  \\ 
Peg~3 & -- & -- & --& $-110^{-16}_{+8}$  & $-16^{-9}_{+10}$ & $-131^{-8}_{+7}$ &   \\  
Cra~2 & -- & --& -- & $108^{-21}_{+36}$  & $-156^{-36}_{+49}$ & $228^{-42}_{+16}$ &  \\  
Dra~2 & -- & -- & -- & $-189^{-16}_{+5}$ & $104^{-36}_{+17}$ & $-265^{-14}_{+4}$ & $-185.1$ \\ 
\hline
\hline
\end{tabular}
\label{TabPropMot2}
\end{table*}
\end{center}
%

\bibliography{master}

\end{document}